Corresponding author: saeed-ul-hassan@itu.edu.pk

# Exploiting Tweet Sentiments in Altmetrics Large-Scale Data

Saeed-Ul Hassan, [a*] Naif Radi Aljohani, [b] Usman Iqbal Tarar, [a] Iqra Safder, [a] Raheem Sarwar, [c]

Salem Alelyani, [d, e] Raheel Nawaz [f]

[a] Department of Computer Science, Information Technology University, 346-B, Ferozepur Road, Lahore, Pakistan.

[b] Faculty of Computing and Information Technology, King Abdulaziz University, Jeddah, Kingdom of Saudi Arabia.

[c] Research Institute of Information and Language Processing, University of Wolverhampton, Wolverhampton, United Kingdom.

[d] Center for Artificial Intelligence (CAI), King Khalid University, P.O. Box 9004, Abha 61413, Saudi Arabia.

[e] College of Computer Science, King Khalid University, P.O. Box 9004, Abha 61413, Saudi Arabia.

[f] Department of Operations, Technology, Events and Hospitality Management, Manchester Metropolitan University, Manchester, United Kingdom.

*E-mail address of the corresponding author: saeed-ul-hassan@itu.edu.pk

**Abstract:** This article aims to exploit social exchanges on scientific literature, specifically tweets, to analyse social media users' sentiments towards publications within a research field. First, we employ the SentiStrength tool, extended with newly created lexicon terms, to classify the sentiments of 6,482,260 tweets associated with 1,083,535 publications provided by Altmetric.com. Then, we propose harmonic means-based statistical measures to generate a specialized lexicon, using positive and negative sentiment scores and frequency metrics. Next, we adopt a novel article-level summarization approach to domain-level sentiment analysis to gauge the opinion of social media users on Twitter about the scientific literature. Last, we propose and employ an aspect-based analytical approach to mine users' expressions relating to various aspects of the article, such as tweets on its title, abstract, methodology, conclusion, or results section. We show that research communities exhibit dissimilar sentiments towards their respective fields. The analysis of the field-wise distribution of article aspects shows that in Medicine, Economics, Business & Decision Sciences, tweet aspects are focused on the results section. In contrast, Physics & Astronomy, Materials Sciences, and Computer Science these aspects are focused on the methodology section. Overall, the study helps us to understand the sentiments of online social exchanges of the scientific community on scientific literature. Specifically, such a fine-grained analysis may help research communities in improving their social media exchanges about the scientific articles to disseminate their scientific findings effectively and to further increase their societal impact.

**Corresponding author:** saeed-ul-hassan@itu.edu.pk

# 1. Introduction

Traditionally, research impact has used citation as the main indicator of research's standing; however, it takes years to see any measurable impact. On the other hand, researchers are increasingly going online to find and share information about science, as well as; have been urged to consider how they can use social media platforms to engage with each other (Hellsten et al., 2019). With the increased usage of social media platforms for scholarly communications, altmetric data are of enhanced interest as it captures realtime scholarly communication data from online platforms (e,g, Twitter, and Facebook) and may be used as an early measure of the research impact.

Altmetrics is the collective domain of social media platforms such as Twitter,[1] Facebook,[2] CiteULike[3] , and MendeleyReadership[4] in relation to research articles to provide metrics on their research impact (Bornmann, Haunschild & Adams, 2019; Drongstrup, Malik & Hassan, 2019; Bornmann, 2014). Among several platforms, Twitter is widely used by scholars to share their opinions concerning research articles (Priem et al., 2011). Recent studies show that tweet sentiments can help predict the early impact of the research articles. Specifically, the papers cited in positive and neutral tweets have a greater impact than those not cited or cited in a negative tweet. However, there is still a need to investigate tweeter data to analyse user sentiments relating to research articles in specific fields. Such a fine-grained investigation is required to fully utilize the findings of existing studies. Specifically, we answer the following research questions in this paper:

1. What is the difference between research communities of different domains regarding tweets containing positive, negative, and neutral sentiments?
2. Does a specific research community is inclined towards a specific aspect of the articles such as methodology, or conclusions?

As mentioned earlier that this article presents a quantitative study to exploit tweet data to analyse user sentiments relating to different aspects of research articles in specific fields. This study helps us to understand the sentiments of online social exchanges of the scientific community on scientific

---

[1] https://twitter.com
[2] https://www.facebook.com
[3] http://www.citeulike.org
[4] https://www.mendeley.com

**Corresponding author:** saeed-ul-hassan@itu.edu.pk

literature, specifically the sentiment of tweets, for better visibility and qualitative assessment of these interesting big data of altmetrics. We identify the sentiment of research communities with respect to their respective fields and to conduct an aspect-based analysis of user expressions related to their research articles. Such a fine-grained analysis may help research communities in improving their social media exchanges about the scientific articles to disseminate their scientific findings effectively and increase their impact.

The following are the three main contributions of the study:

- Lexicon generation: We design a harmonic means-based statistical measure to generate a specialized lexicon to conduct this investigation, which helps improve the performance of the sentiment analysis task. This is because general sentiment lexicons calculate the sentiment tendency of a word without considering domain knowledge. However, the sentiment contained in just a few words is inevitably domain-dependent. Therefore, generic sentiment lexicons used by SentiStrength report poor performance in various applications. For this reason, in this investigation, we design a new measure to generate a new lexicon for our altmetrics data to determine both domain-specific and expressive terms and then feed it to SentiStrength to identify the sentiments of the tweets. Specifically, we computed the rate and frequency metrics of each term or 'token.' Next, we compiled statistical measures, such as the harmonic mean, using a cumulative distributive function for both positive and negative terms. The resulting descending-order list of lexicon terms shows the most meaningful and domain-dependent tokens in sentiment expressions and provides meaningful insights into the terms used in opinion mining in this altmetrics domain.

- Based on our newly generated lexicon, we designed a threshold-based mechanism to compute domain-wise article-level sentiment. We found that research communities exhibit dissimilar sentiment towards their respective fields.

- We design a method to perform an aspect-based analysis of user expressions related to the research article, such as its title, abstract, methodology, conclusion, and results. We found that research communities focus on different aspects of an article. Researchers in fields such as Medicine and Economics, Business & Decision Sciences show more interest in an article's findings than its title, abstract or methodology. Interestingly, fields such as Engineering and Computer Sciences address more the techniques designed. Likewise, in Health Professions & Nursing, the scientific community primarily discusses articles on the basis of both their abstract and their findings.

The structure of the rest of this article is as follows: section 2 includes previous research work, associated concepts and a literature review. In section 3, we present the details of our dataset, followed by a discussion on our approach to lexicon creation and tweet sentiment analysis. Section 4 presents our data and insights on the results achieved. We end this study with concluding remarks.

## 2. Literature Review

Altmetrics has a very broad scope, and many studies have been undertaken to define the extent of the term, the type of research measures that it may or may not provide, and whether there is enough data to indicate any impact (Priem et al., 2010). Haustein et al. (2016) regard altmetrics as an umbrella term for an article-level metrics of research impact that encompasses several social media platforms, such as Twitter, Facebook, MendeleyReaders, CiteULike, Google+ (El Rahman, AlOtaibi & AlShehri, 2019). Altmetrics data are increasing all the time, and multiple organizations gather them, including altmetric.com, Impact Story, and Plum Analytics. These organizations collect all online activities concerning research articles and offer these data for research purposes. We have observed a promising increase in research into sentiment analysis and opinion mining of altmetric dataset of researchers, publishers, universities, and funders in the past few years, hence there is a growing demand for standards and new challenges to ensure best practice (Vairetti et al., 2020; El Rahman, Al Otaibi & AlShehri, 2019; Wouters, Zahedi & Costas, 2019). In the following subsections, we provide a brief overview of previous studies to highlight the quality and challenges of our altmetrics dataset and approaches that we used in sentiment analysis of Twitter altmetrics.

### 2.1 A Brief Review of Altmetrics

Researchers and academics are increasingly using online research tools to access, download, bookmark, recommend, discuss, share, and evaluate ongoing research. Through their online

Corresponding author: saeed-ul-hassan@itu.edu.pk

presence, they are creating huge volumes of online data that can be used in altmetrics. Traditionally, the relevance and actual impact of a research article have been gauged by its citation count, but this has the inherent problem of being sluggish. The use of this conventional citation metrics may be superseded by mining altmetrics data, as this can produce useful insights (Diamantini et al., 2019; Gallo et al., 2019; Haustein et al. 2014; Hammarfelt, 2014; Priem et al., 2012; Thelwall et al., 2013).

Of the altmetrics indicators such as Facebook, Google+, CiteULike, Mendeley, Wikipedia, and other online blogs, Twitter is the platform most used social by scholars and researchers, and many studies have investigated this use. Priem and Costello's study (2010) investigated 46,515 tweets from a sample of 28 scholars and examined their attitudes and practice to Twitter for scholarly discussion. It explored how often they tweeted research articles, and the results revealed that, while they use it in this way, such citation is different from the traditional citation. The study concluded that Twitter citations are much more rapid and that Twitter does indeed have an impact on scientific research. To find any common pattern of use among the disciplines or whether they are clearly different, Holmberg and Vainio (2018) performed a cross-disciplinary analysis on how and why researchers use Twitter. They analysed 10 diverse disciplines and categorized the tweets of selected scholars as: Scholarly communication; Discipline-relevant; Not clear; and Not about science. Their results show a clear difference in Twitter usage between scholars from these various disciplines. Priem and Costello's study (2010) discusses the quantity and quality of altmetrics data that are generated over the years. As well as citation metrics, the authors correlated article-level metrics on various altmetrics platforms. They answered the main question, whether it can predict citation counts and is indeed an early measure of research impact, as their comparison of altmetrics and traditional citation revealed its significant contribution to the early prediction of citations. However, they concluded that altmetrics is different from citation count, as the impact that is captured is across a highly varied audience, which may suggest a much wider societal impact in multiple educational, cultural, environmental, and economic fields. Haustein (2019) discussed both how social media signals are revealed in various scientific fields and that they differ by document type. The results indicate that, in general, mentions of research articles on online platforms are somewhat low; however, Twitter has the best coverage of all social media platforms. The study also explored which altmetrics indicators have the most significant connection to citation count, and concluded that Twitter and online blogging have the best correlation with traditional metrics. Further analysis by Haustein (2019) showed that shorter documents, such as editorials, news articles and letters, tend to receive more online coverage than longer, more complex documents.

## 2.2 Tweet Sentiments of Altmetrics

Sentiment analysis algorithms either rely on machine learning or lexical methods. The machine learning methods partition text into words or word n-grams, learn which of these features are associated with sentiments based on human coded text, and use this information to predict the sentiment of the test sample. On the other hand, lexical methods use a list of sentiment words and their polarities with grammar structure knowledge such as a negation role to predict the sentiment of the text. Nevertheless, lexical methods report better accuracy for social media texts and are less likely to choose indirect indicators of sentiment that generate spurious sentiment patters. For instance, machine learning methods may choose unpopular politicians' names as negative features since they tend to occur in the negative text (Chaturvedi et al, 2018; Pak et al., 2010; Agarwal et al., 2011). Typically, people use shortened forms of words and emoticons when writing on social web platforms, which increases the need to create tools to identify feelings in a short text (Plaza-del-Arco et al., 2019; Li et al., 2010). Thelwall, Buckley and Paltoglou (2012) devised an algorithm, 'SentiStrength', that works in both supervised and unsupervised cases. It adopts a lexical approach in which a list of terms is tagged with positive or negative sentiments on a scale of -5 to +5 and, on the basis of the occurrence of these terms, it predicts the sentiment of a text. The lexicon model may include additional information, such as emoticon lists and semantic rules for dealing with negation words. The SentiStrength algorithm shows good results when performing sentiment analysis on the datasets of web networks (Myspace, Twitter, Facebook, YouTube, BBC Forums). It works well with social web data for which no training dataset is available to detect sentiment, thus are recommended for applications in which direct, effective terms are exploited by performing sentiment analysis.

Exploiting the tweet sentiment in Altmetrics data requires a sentiment analysis tool that performs well on social media text which is generally short and contains non-textual elements such as emoticons and is categorized as non-standard expressive text. In addition to this, it requires a sentiment analysis tool that can determine the positive and negative sentiments simultaneously.

This is because psychological research reports that humans can experience negative and positive emotions simultaneously. Furthermore, it requires a sentiment analysis tool that works well with low or no training data. This is because, for some fields, there is less amount of altmetrics data available for analysis. Unlike machine learning-based sentiment analysis tools, the SentiStrenth tools, which is a lexical method have all these properties.

Scholars frequently use Twitter as a discussion platform to share their opinions on research. Perhaps, for this reason, digital libraries and journal websites are increasingly using tweet counts as a measure of the impact of research (Fernando et al., 2019). To evaluate its use as an alternative measure of impact, Zimbara et al. (2018) raise the need to analyse the opinion contained expressed in tweets about articles. Shema, Bar-Ilan and Thelwall (2015) performed a pilot study on 270 randomly collected tweets about research articles and analysed the kind of opinion that they expressed on the articles, and whether, as a research impact measure, the proportion of negative tweets to the overall tweet counts can be ignored. Their results showed that tweets about scholarly articles are mostly objective, consisting of either the title or points from a brief summary. They concluded that, as tweets about an article contain little sentiment or opinion on the article, they cannot provide much insight into its research. Friedrich et al. (2015) analysed the tweets of articles and reviews published in 2012 in WoS, as captured by altmetric.com. The dataset consisted of 487,610 tweets, mentioning 192,832 articles. The results showed that 11.0% contain positive sentiments and 7.3% negative, and 81.7% are neutral. Disciplinary analysis shows that fields such as Psychology, the Humanities and Social Sciences contain the most sentiment in their tweets, while fields such as Physics, Chemistry and Engineering express the least (Yu, 2017). Hassan et al. (2019b) state that the Twitter-user influence score is a highly important feature in the classification of highly cited articles.

Additionally, to ascertain scholarly impact through altmetrics events there are challenges to be addressed. Studies have provided evidence that it is not actually the scientific merits or characteristics of an article that is captured by online or social media attention. Haustein (2016) showed that a curious or humorous article receives more tweets. Furthermore, Zheng et al. (2017) showed that scientific journals may use social media as a platform for their promotional campaigns, creating an enhanced level of altmetrics events about certain research. A study by Robinson-Garcia et al. (2017) pointed out that usage of scholarly online and social platforms is almost devoid of sentiment and, in most cases, it offers no opinion. However, citation presents the same issue: a study by Didegah et al. (2018) revealed that the intentions behind creating a citation vary, and some actually relate to something other than the research itself. Friedrich et al. (2015) observed that long abstracts of medically related articles receive more citations, whereas longer titles in Psychology receive fewer.

Since many studies have explored techniques of sentiment analysis, certain aspects of citations using altmetrics data show a marked variation, aside from their scientific merit and approach. Most measurement of the sentiment and opinion of the people tweeting about research has been carried out quantitatively. Our study takes a more qualitative approach, exploiting tweet sentiment and opinion mining at a higher level, using document-level sentiment analysis and aspect-based sentiment analysis. This qualitative content analysis could introduce new viewpoints to altmetrics research.

## 3. Dataset and Methodology

In this section, we discuss the proposed method. The proposed method consists of five parts: altmetrics data collection, tweet pre-processing, lexicon generation, combining article-level tweets, and analysis (see Figure 1). Each part of the proposed method is explained in the following subsections.

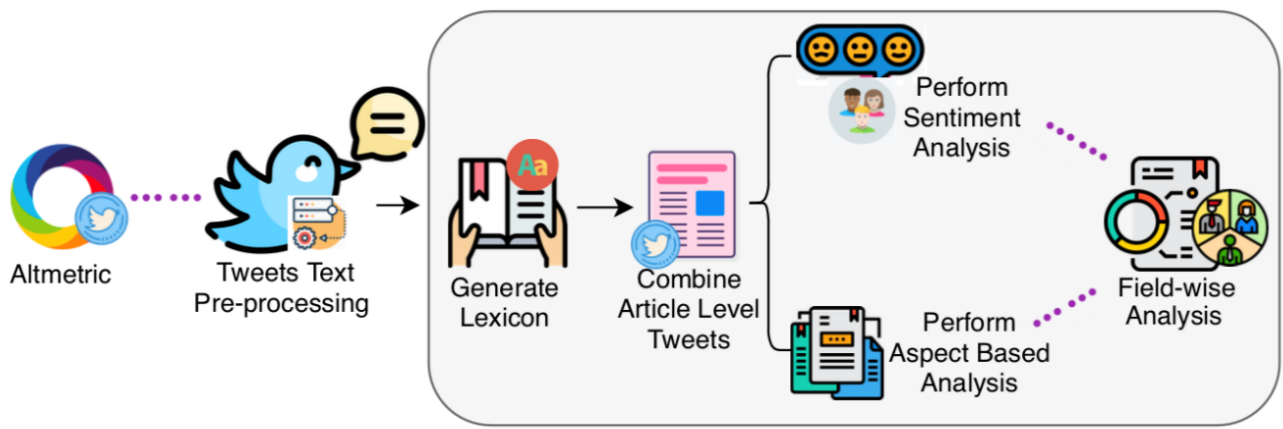

**Figure 1:** Detailed architecture of proposed methodology

**Corresponding author:** saeed-ul-hassan@itu.edu.pk

## 3.1 Dataset

The corpus comprised altmetrics data collected by Altmetric.com from July 2011 to June 2016. Note that altmetrics.com[5] is the most important collector of social media content, offering this data for research purposes. The database consists of aggregated content from online platforms such as Twitter, Google+, Facebook, CiteULike. Twitter is the chief contributor. From the altmetric data, we extracted 1,083,535 research articles that each had at least one citation and one tweet. While using the tweet URL, we fetched 6,482,260 tweets from Twitter, we retrieved the articles' citation count using Scopus API along with the disciplinary information provided by the Scopus subject-category scheme.

For cross-disciplinary analysis, the dataset was divided into scholarly disciplines by the ASJC subject classification scheme. Inspired by the work of Haddawy et al. (2017), the top-level ASJC disciplines were merged into 16 disciplines by combining Agricultural, Biological Sciences & Veterinary; Biochemistry, Genetics & Molecular Biology; Chemistry; Computer Science; Earth Planetary Sciences; Engineering; Environmental Science; Economics, Business & Decision Sciences; General; Material Science; Health Professions & Nursing; Mathematics; Medicine & Medical Sciences; Physics & Astronomy; Social Sciences; Other Life and Health Sciences.

## 3.2 Pre-processing

To demonstrate the need for pre-processing, Table 1 shows a few examples of the unprocessed tweet text. To obtain the clean text for lexicon creation and sentiment analysis, we performed the following pre-processing steps: (i) we detected and removed all non-English tweets; (ii) since tweet text sometimes contains research-specific terms taken from the article's title that are not actual opinion specific to the research article, we removed any such terms to avoid false allocation of sentiment (Robinson-Garcia et al., 2017); (iii) we used Beautiful Soup Python Library[6] to decode HTML encoding, such as '&', '"', and so on, into general text; (iv) we removed tags like '@mention' from the tweet text using regular expressions (REs) '(r' @[A-Za-z0-9]+')'; (v) we removed URLs using REs (r' https?://[A-Za-z0-9./]+' and r' www.[^ ]+'); (vi) we found and removed any Unicode Transformation Format(UTF)-8 encoding patterns of characters' \xef\xbf\xbd' using UTF decoding; (vii) we kept numbers as text, only removing the '#' character using REs ('[^a-zA-Z]'; (viii) we dropped any duplicate and null-text tweets; (ix) we carefully handled negation words to avoid their destruction in pre-processing by preparing a list of common negations (words with apostrophic combination), such as isn't (is not), aren't (are not), wasn't (was not), weren't (were not), haven't (have not), hasn't (has not), couldn't (could not), shouldn't (should not), and so on, converting them into two words; (x) last, we removed unnecessary blank spaces, performed tokenization and lowercasing, and rejoined tokens to form proper sentences.

---

[6] https://pypi.org/project/beautifulsoup4/

**Table 1**: Diversity of tweet texts in the altmetrics dataset

| Altmetric_ID | Tweet_ID | Text |
|---|---|---|
| 786919 | 2.1642E+17 | RT @ohsuneuro: Personalized medicine comes to DBS - tailoring freq of stim based on pt intrinsic STN oscillations http://t.co/jPx54P1e... |
| 786922 | 2.9399E+17 | Risk of fractures in #MS patients. Worth looking into. We all need to be careful. http://t.co/JYFliqVR #GavinGiovannoni #SecureACure4MS |
| 787090 | 2.1233E+17 | New paper from Professor Brendan Kennedy. *Physical Review* B, 85(17), 174110, 2012. http://t.co/TP1XWM37 |
| 1822747 | 4.0998E+17 | RT @richardheinberg: important new peer-reviewed meta-study on peak oi. http://t.co/UBjWo6rgOJ |
| 1822815 | 4.1388E+17 | RT @CaloriesProper: Designing future prebiotic fiber to target the metabolic syndrome. http://t.co/H4cOxvaJMY #galactooligosaccharides |
| 1822863 | 3.894E+17 | 'Leaf mesophyll cond. and leaf hydraulic cond.: an intro to their measurement and coordination.' http://t.co/FfrZzUgvB1 @JXBot #plantphys |

**Corresponding author:** saeed-ul-hassan@itu.edu.pk

### 3.3 SentiStrength

Exploiting the tweet sentiment in altmetrics data requires a sentiment analysis tool that performs well on social media text which is generally short and contains non-textual elements such as emoticons and is categorized as non-standard expressive text. In addition to this, it requires a sentiment analysis tool that can determine the positive and negative sentiments simultaneously. This is because psychological research reports that humans can experience negative and positive emotions simultaneously. Furthermore, it requires a sentiment analysis tool that works well with low or no training data. This is because, for some fields, there is less amount of Altmetrics data available for analysis. Unlike machine learning-based sentiment analysis tools, the SentiStrength tools, which is a lexical method have all these properties. SentiStrength uses a lexical approach to identify the sentiments of social media texts. Specifically, it simultaneously determines the strength of positive (on a scale of 1 to 5) and negative (on a scale of -1 to -5) emotions because of psychological research reports that humans can experience negative and positive emotions simultaneously. The SentiStrength sentiment detection algorithm was initially developed on a sample set of 2,600 MySpace classifications used for the initial testing. The key elements of the SentiStrength are described below.

*Sentiment Word Strength List:* The essence of the SentiStrength algorithm is the word strength list for the sentiment that contains 298 positive terms and 465 negative terms. These terms are classified for either positive or negative sentiment strength with a score from 2 to 5. During the development stage, the sentiment strength score was given based upon the human judgment but later while training, scores are modified using automatic mechanisms (details are given below).

*Optimize the sentiment word strengths:* The algorithm begins with the baseline strengths for the *Sentiment Wordlist*, given by human judgment. Then it determines for each term if an increase or decrease in intensity (strength score) by 1 will either improve the accuracy of the classifications or not. Any adjustment that improves the overall accuracy by at least 2 is retained.

*Spell Correction Algorithm: The a*lgorithm finds the regular word spellings that are misspelled by the use of repeated letters. For instance, the word "Helllllloooo" would be identified by this algorithm as "hello." The algorithm (a) removes repeated letters above twice (e.g., helllo-> hello);

**Corresponding author:** saeed-ul-hassan@itu.edu.pk

(b) removes repeated letters appear twice for letters rarely occurring twice in English (e.g., niice-> nice).

*The Booster Words:* It involves those words that bring a raise or decrease in the emotion of the words that follow, whether positive or negative. For instance, extremely, so, very, much, immensely, etc. are those words the boost the emotion of the subsequent words, whereas, some, few, etc. words cause a decrease in the words' emotions. Similarly, an *inverted emotions words list* contains those words that invert the emotions of the successive word. For instance, if the words "so happy" have positive strength score 4 then "not so happy" should have a negative score of 4.

Furthermore, the SentiStrength algorithm takes *exclamation marks* into account as well and gives a minimum of positive strength 2 to all sentences with exclamation marks. Also, in the case of *repeated punctuations*, a sentence having at least one exclamation mark brings a force boost of 1 to the preceding emotion word.

SentiStrength algorithm was also compared and evaluated with the standard ML algorithm such as SVM, SVM regression, simple logistic regression, JRip rule-based Classifier, Decision Table, J48 classification Tree, Naïve Bayes, Multiple perceptron and AdaBoost using Weka. These ML algorithms use frequencies of each word present in the sentiment word list as a simple feature set and an extended feature list that takes all text elements into account, similar to SentiStrength. It works well with the non-standard expressive text. Moreover, it works well without training data on several social media texts and reports human-level accuracy in most cases.

### 3.4 Lexicon Creation of Tweets in Altmetrics

By adapting to the SentiStrength tool for sentiment analysis of Altmetrics data our whole data is tagged into positive, negative, and neutral sentiments. But since SentiStrength is a generic lexicon-based tool and the use of words varies a lot from topic to topic. Therefore, generic sentiment lexicons used by SentiStrength report poor performance in various applications. Thus, we propose an improved scoring method for our Altmetrics corpus so as to see the most expressive terms of opinion for both positive and negative sentiments. Specifically, we design a harmonic means-based statistical measures to generate a specialized lexicon to conduct this investigation which help

Corresponding author: saeed-ul-hassan@itu.edu.pk

improve the performance of the sentiment analysis task. More specifically, we design a new measure to generate a new lexicon for our altmetrics data to determine both domain-specific and expressive terms and then feed it to SentiStrength to identify the sentiments of the tweets (see Algorithm 1).

We extracted 152,673 words/features from our dataset, along with their positive, negative and total occurrence scores, using the Python Count Vectorizer[7] method. Our intuition is that if a word appears more frequently in positive class as compared to a negative one, then it should be more characterized by a positive term. Similarly, if a word appears more frequently in negative class as compared to the positive one, then it should be more characterized by negative terms. Thus, for each term in our dataset, we calculated Positive Rate (PR) and Negative Rate (NR). The PR of a term is calculated as the ratio of the relative frequency of the term in positively identified texts to the frequency of the term in all texts (see Equation 1), while its NR is the ratio of the relative frequency of the term in negatively identified texts to the total frequency of the term in all texts (see Equation 2).

$$PR = \frac{Positive\ frequency}{Total\ Frequency} \qquad (1)$$

$$NR = \frac{Negative\ frequency}{Total\ Frequency} \qquad (2)$$

We then sorted the terms by the rates and found no meaningful pattern in the top-scoring terms. Specifically, we found that words with the highest positive rate have zero frequency in negative tweets, but the overall frequency of these words is too low to consider it as a guideline. Next, we ascertained the rate of occurrence within a class by calculating Positive Frequency (PF) and Negative Frequency (NF) metrics, as shown in Equations 3 and 4. This new metric resulted in almost the same ranking as the original term frequencies.

$$PF = \frac{Positive\ Frequency}{Sum\ of\ positive\ frequencies} \qquad (3)$$

---

$$NF = \frac{\text{Negative Frequency}}{Sum\ of\ negative\ frequencies} \qquad (4)$$

Since our intuition is to rank terms in order of their positive sentiment value, so we generate the cumulative distribution function (CDF) values of PR and PF for the positive sentiment value; and CDF values of NR and NF for the negative sentiment values. CDF is a probability distribution function of X that is evaluated at x, and it measures the probability that X will take a value less than or equal to x, as shown in Equation 5:

$$F(x) = P(X \leq x) \qquad (5)$$

The calculation of CDF value of PR or PF provides insight into their ranks in the distribution. Next, we combine CDF of PR and CDF of PF together to produce a metric that has a reflection of both PR and PF. That is CDF help find terms' associations using their rate and frequency values. For instance, the term 'Excellent' scored 0.83 CDF of PR value and 1.00 CDF of PF. This means that roughly 83% of tokens will have a PR value of less than or equal to 0.99 and, for PF, all have a PF value of less than or equal to 0.001786. The CDF is used here to give the cumulative values of the distribution of PR and PF.

Next, we combine PR-CDF and PF-CDF together to produce a metric that has a reflection of both PR and the PF. Upon looking at the values, we found that the PR-CDF spans from 0 to 1 and the PF-CDF values are distributed in a smaller range, i.e., 0 to 0.4. Consequently, taking the arithmetic average of these two numbers will dominate the PR over the PF value, thus instead we rely on the harmonic mean. Finally, we computed the harmonic mean (HM) of the CDF values for both the rate and frequency metrics. HM is the reciprocal of the arithmetic mean of that reciprocal. It is appropriate to use the harmonic mean when the metrics include outliers that, which could skew the results. Equations 6 and 7 show the HM for positive (HMP) and negative (HMN) terms, respectively, while $n$ represents the number of metrics:

$$HMP = \frac{n}{\frac{1}{PR}+\frac{1}{PF}} = \frac{n\,(PR\,.PF)}{PR+PF} \qquad (6)$$

$$HMN = \frac{n}{\frac{1}{NR}+\frac{1}{NF}} = \frac{n\,(\,PR\,.PF)}{NR+NF} \qquad (7)$$

It is important to note that HM works the same as the F-score in terms of precision and recall metrics. Therefore, HM supports a cumulative score for all terms, providing a useful scoring mechanism for our tokens, as the descending order list shows the most meaningful and domain-dependent tokens in sentiment expressions. Appendix A (Tables A1 and A2) lists the top-100 positive and negative words in our altmetrics dataset.

---

**Algorithm 1** Lexicon Creation of Tweets

1: **procedure** LexiconCreation (Altmetrics Dataset)
2:     Words ← CountVectorizer(Altmetrics Dataset)
3:     **for** each term in Words list **do**
4:         Calculate PR ← $\dfrac{+Frequency}{TotalFrequency}$ and PF ← $\dfrac{+Frequency}{\sum Freq(+Tweets)}$
5:         Calculate NR ← $\dfrac{-Frequency}{TotalFrequency}$ and NF ← $\dfrac{-Frequency}{\sum Freq(-Tweets)}$
6:         Compute $PR\_CDF \leftarrow P(PR \leq x)$ where CDF is a probability distribution function of PR that is evaluated at x
7:         Compute $PF\_CDF \leftarrow P(PF \leq x)$
8:         Find $Pos\_Score \leftarrow \dfrac{n(PR\_CDF * PF\_CDF)}{PR\_CDF + PF\_CDF}$
9:         Calculate $NR\_CDF \leftarrow P(NR \leq x)$
10:        Compute $NF\_CDF \leftarrow P(NF \leq x)$
11:        Find $Neg\_Score \leftarrow \dfrac{n(NR\_CDF * NF\_CDF)}{NR\_CDF + NF\_CDF}$
12:    **end for**
13:    **return** $Pos\_Score, Neg\_Score$
14: **end procedure**

---

### 3.5 Article-Level Sentiments of Tweets in Altmetrics

To analyse the article-level sentiment of our altmetrics data on the basis of their Alt_ID, we combined the tweets about each article with at least 30 tweets. The objective of this level of analysis is to express a single sentiment for the whole article, and it assumes that all the sentences within a document refer to a single entity. We had a total of 61,233 distinct Alt_IDs for each research article with at least 30 tweets, and we computed sentiment scores for each using our newly created lexicon in SentiStrength. Once the scores were applied to the positive and negative terms, we achieved an average that ranged from 0.0 to 1.0. We referred to those values above 0.7 as positive and those below 0.3 as negative, and scores between the two as neutral.

Corresponding author: saeed-ul-hassan@itu.edu.pk

**Table 2**: Article section and keyword

| Title | Title, subject, topic |
|---|---|
| **Abstract** | Abstract, overview, summary |
| **Methodology** | Method, material, calculation, procedure, tool, approach, model, technique, experiment |
| **Results & Conclusion** | Result, evaluation, conclusion, value, discussing, showing, finding |

### 3.6 Aspect Analysis of Tweets in Altmetrics

Liu and Fang (2017) define an opinion as a quintuple: ($e_i$, $a_{ij}$, $h_k$, $t_l$). Here, $e_i$ and $a_{ij}$ together represent the opinion target, where $e_i$ is the entity as the main target of opinion, $a_{ij}$ is an aspect of entity $e_i$ for which opinion is being generated, $h_k$ is the opinion holder and $t_l$ is the time when the opinion is expressed by $h_k$.

Using the above definitions, we performed domain-wise, article-level, aspect-based analysis of our altmetrics data. In this instance, the entity was a research article and the aspects were the title of the article, its abstract, methodology and the conclusion discussed at the end. The objective was to gauge community behaviour in tweeting about an article, by domain. First, on the basis of their Alt_IDs and domain code (QRR_IDs), we combined all tweets about each article with at least 30 tweets. Note that an article may fall into multiple domains, so the combined sum of articles (Alt_IDs) was 153,336. Next, using the double-counting method, we identified the various aspects of an article that were expressed by researchers in their tweets, as typically stated in the keywords, as shown in Table 2. For every tweet in which the opinion referred to the entire article, that opinion was marked as a general aspect of the article.

## 4 Experimental Results

In this section, we discuss the results obtained by our various analysis techniques, along with their significance to the different domains.

### 4.1 Distribution of Tweet Sentiments

Using SentiStrength with the domain- and emotion-specific terms, as prescribed by Hassan et al. (2019a), we classified as positive, negative and neutral a total of 6,482,260 tweets, relating to 1,083,535 altmetrics documents. We found that 22% were positive, around 14% negative and 64% neutral, as shown in Figure 2. Furthermore, we explored our altmetrics tweets dataset to detect any significant change in behavior in the usage of tweet sentiment. Figure 2 illustrates that there was no significant increase in tweet sentiment during the period 2012 to 2016.

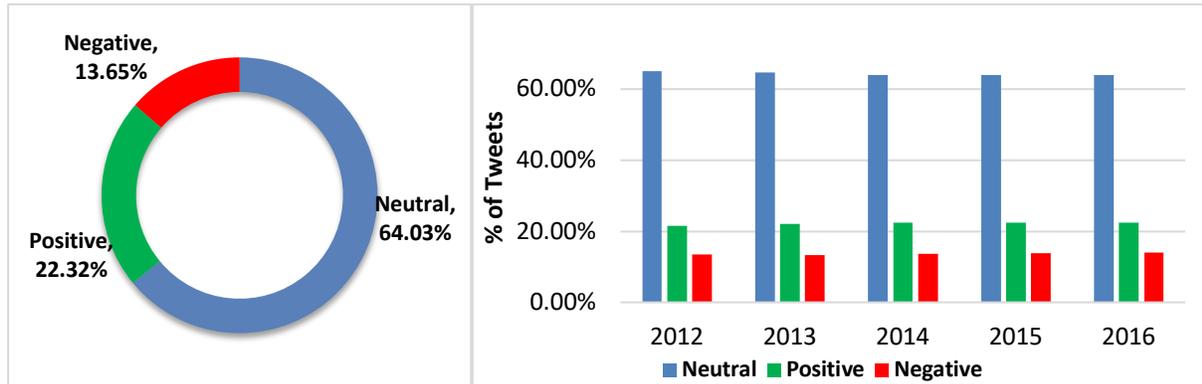

**Figure 2:** Distribution of tweet sentiment

## 4.2 Lexicon for Altmetrics Data for Sentiment Scores

SentiStrength is a generic lexicon-based tool with a generic text corpus. Since each text corpus is different in its nature and the use of words in subjects varies widely, we created a relative scoring technique based on our altmetrics corpus. We calculated the harmonic mean of CDF scores (normCDF_HM) for PR and PF, and NR and NF. The normCDF_HM provides a significant scoring pattern for the corpus unique terms. Table 3 gives the descending list of the most meaningful tokens in our corpus in terms of sentiment expression. Appendix A (Table A1 and A2) contains the top-50 positive and negative tokens in the altmetrics dataset. Figure 3 illustrates the interesting pattern displayed by the normCDF_HM scores for both the rate and frequency metrics. The tokens shown at the top left are more positive, and the ones at the lower right are more negative. In this way we created our own lexicon for the altmetrics corpus, and it will prove useful in the classification of tweets sentiment in future.

**Table 3**: Terms in descending order of positive harmonic mean

| Token | Total count | Positive HM score | Negative HM score | Token | Total count | Positive HM score | Negative HM score |
|-------|-------------|-------------------|-------------------|-------|-------------|-------------------|-------------------|
| Excellent | 6,669 | 0.9077 | 0.2533 | Positive | 4,767 | 0.9008 | 0.2737 |
| Novel | 3,713 | 0.9064 | 0.2558 | Nice | 14,630 | 0.9007 | 0.2886 |
| Amazing | 2,907 | 0.9054 | 0.2578 | Thanks | 6,169 | 0.9006 | 0.2768 |
| Congratulations | 1,930 | 0.9051 | 0.2487 | Special | 3,458 | 0.8997 | 0.2739 |
| Awesome | 3,697 | 0.9040 | 0.2627 | Cool | 9,996 | 0.8991 | 0.2885 |
| Success | 2,319 | 0.9031 | 0.2606 | Interesting | 33,787 | 0.8987 | 0.3086 |
| Wow | 3,296 | 0.9030 | 0.2648 | Greater | 4,340 | 0.8963 | 0.2856 |
| Interested | 4,181 | 0.9024 | 0.2678 | Hope | 2,277 | 0.8955 | 0.2779 |
| Exciting | 2,343 | 0.9023 | 0.2628 | Love | 4,026 | 0.8947 | 0.2892 |
| Great | 25,625 | 0.9013 | 0.2952 | Pretty | 2,171 | 0.8937 | 0.2803 |

## 4.3 Article-Level Summarization for Altmetrics Domains

To perform article-level summarization, using SentiStrength we combined all tweets about an article with at least 30 tweets into a single document and computed the document-level sentiment. Note that articles with fewer tweets were discarded. Of the total of 61,233 unique articles, we found that around 82.55% contained neutral sentiments, followed by 17.35% with positive sentiments and only 0.1% of the articles were negative. The results suggest that, at article-level, the negative sentiments are quite insignificant.

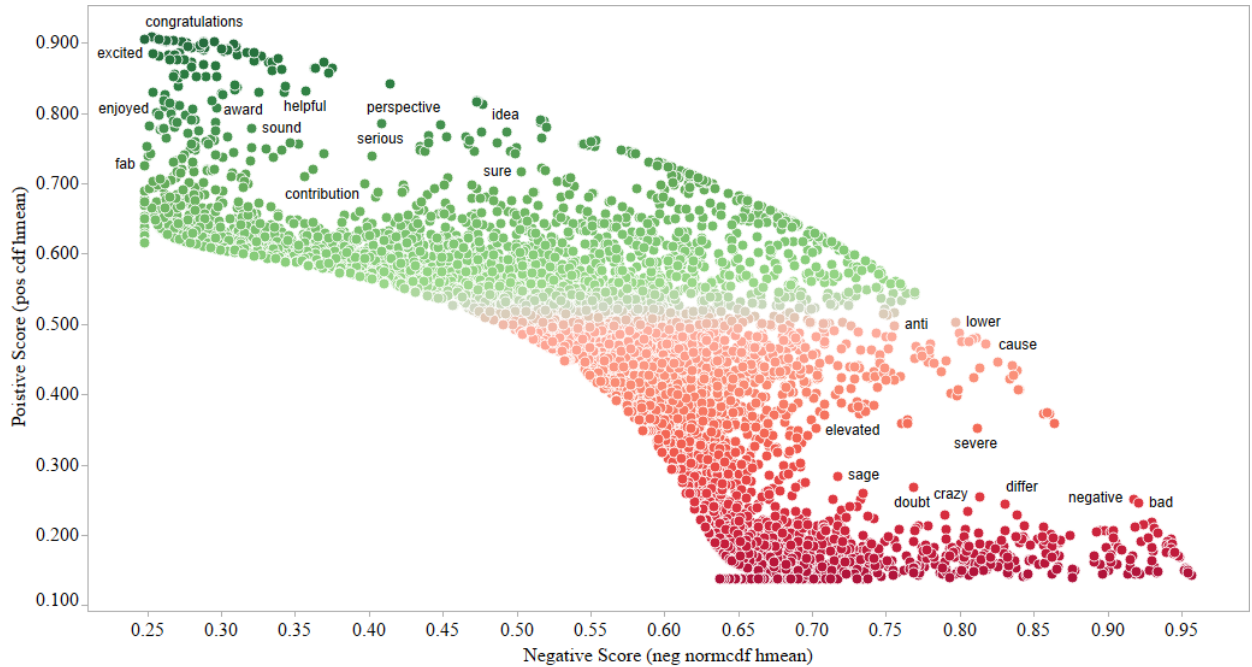

**Figure 3:** Scatter plot of tokens' positive and negative scores

In addition, we performed domain-level sentiment analysis using article-level summarization in order to measure user behaviour across the domains. We aggregated article-level tweet documents on the basis of QRR_Field, and used SentiStrength, enriched by the new lexicons, to calculate the positive, negative and neutral sentiment scores. Table 4 gives a summary of the results, along with the normalized positive and negative sentiment scores from 0 to 1. Since the entity is not supposed to be single, we do not attempt to suggest that domain-level summarization will give an opinion about the domain. Rather, it helps to show the intent and to indicate the behaviour of the users by their domain.

The results show that researchers expressed more positive opinion in domains such as Arts & Humanities, Computer Science and Chemistry, while the fields of Medicine, Health Professions & Nursing and Other Life & Health Sciences attracted more negative opinions from their respective scientific communities. Figure 4 presents a scatter plot to illustrate the community behaviour's in each research domain. With reference to normalized sentiment score (HM Score), the domains expressing more positive opinions are at the top left, while those with a high value of negative sentiment are at lower right.

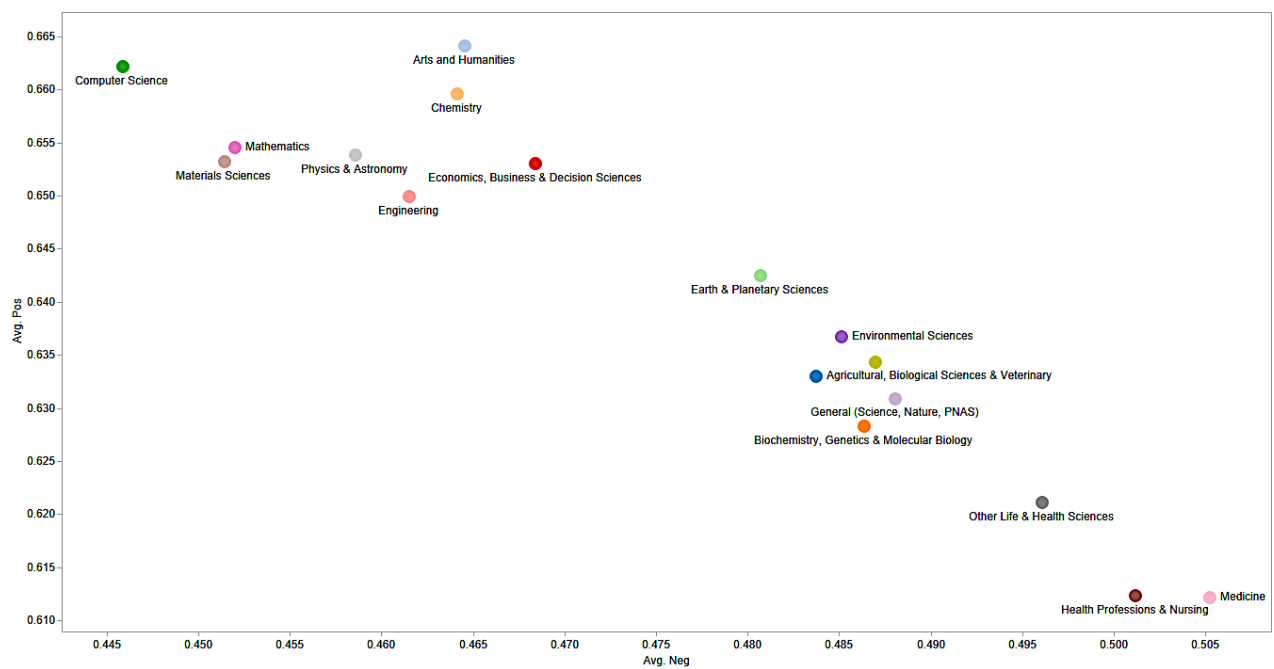

**Figure 4:** Scatter plot of research domains in terms of positive and negative sentiment

Furthermore, we employed distribution analysis to see the difference from a normal distribution of Alt-Domains by fitting the tweet scores to a bell curve, as shown in Figure 5. The results indicate that Twitter usage in the domains of Arts & Humanities, Chemistry, Computer Sciences, Material Sciences and Mathematics are positive, while in Medicine, Health Professions & Nursing and other Life & Health Sciences it is towards the negative. It was found that domains such as Earth & Planetary Sciences are neutral, overall.

## 4.4 Article-level, Aspect-based Analysis

For domain-wise, aspect-based analysis, on the basis of their research domains, we compiled article-level tweet documents for the 61,233 unique articles in our altmetrics dataset that had at least 30 tweets each. Note that an article may fall into multiple domains, so the combined sum of articles is 153,336. Table 4 presents a summary of the results for article-level tweets that contain the users' opinion of the title, the abstract, methodology, and conclusion and results. Note that we created a separate category 'Other', for where a whole article is discussed in general. For articles in our dataset with at least 30 tweets, Table 5 shows the proportion that specifically discusses their various aspects in terms of their respective subject domain.

**Corresponding author:** saeed-ul-hassan@itu.edu.pk

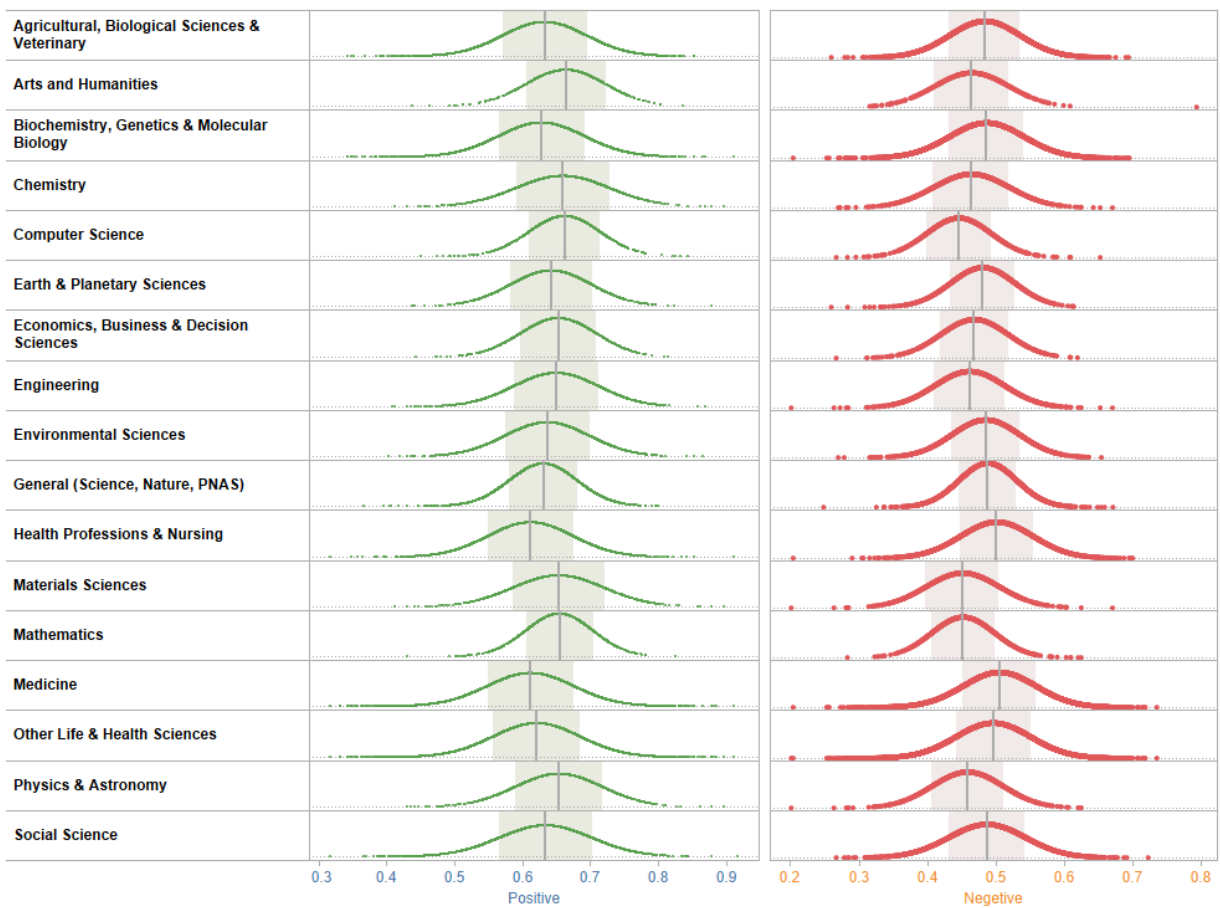

**Figure 5:** Normal distribution of tweet sentiment in various research domains

Regarding using the title in expressing an opinion, across the fields we found that General (*Science, Nature, PNAS*) was prominent, with 2.47% articles being debated in this way, followed by Arts & Humanities and Social Sciences, with 2.29% and 2.09% articles respectively. In terms of abstract-based opinion, the domain of Health Professions & Nursing is significant, with 5.39% articles debated on this basis, followed by Arts & Humanities at 5.20% and Computer Sciences at 4.98%. We noted that in Material Sciences, 12.04% of article tweets concentrated on the article's methodology, and in Physics & Astronomy and Chemistry this was over 8%. Interestingly, it is important to note that researchers in fields such as Engineering and Computer Sciences address the techniques designed relatively more. Furthermore, in terms of debating an article on the basis of its results and conclusions aspects, we found that the domain of Economics, Business & Decision Sciences was the most notable of all domains, at 11.75%. Similarly, this domain (11.75%), Medicine (10.84%), that of Health Professions & Nursing (10.11%) and General

**Corresponding author:** saeed-ul-hassan@itu.edu.pk

(*Science, Nature, PNAS*) (10.03%) appear to be most concerned to address aspects of articles' results and conclusions.

By contrast, analysis of the field-wise distribution of an article's aspects shows that in Medicine and Economics, Business & Decision Sciences, researchers show more interest in the findings than the title, abstract or methodology. Likewise, the Health Professions & Nursing scientific community primarily discusses articles' abstracts and findings. Those in General (*Science, Nature, PNAS*) are more focused on an article's title and research results than are other fields. In the case of Material Sciences, 12% of all articles are debated on the basis of their methodology. This clearly indicates that this community is much concerned with the methods that are designed and presented by an article. Overall, the analysis suggests that researchers appear to be descriptive when exploring the various aspects of an article.

**Table 4**: Summary of domain-level sentiment analysis results

|  | # of documents with at least 30 tweets | Avg. positive sentiment score | Avg. negative sentiment score |
|---|---|---|---|
| Arts & Humanities | 468 | 0.66 | 0.46 |
| Computer Science | 979 | 0.66 | 0.45 |
| Chemistry | 1414 | 0.66 | 0.46 |
| Mathematics | 950 | 0.65 | 0.45 |
| Physics & Astronomy | 1208 | 0.65 | 0.46 |
| Materials Sciences | 1094 | 0.65 | 0.45 |
| Economics, Business & Decision Sciences | 924 | 0.65 | 0.47 |
| Engineering | 2491 | 0.65 | 0.46 |
| Earth & Planetary Sciences | 1438 | 0.64 | 0.48 |
| Environmental Sciences | 2753 | 0.64 | 0.49 |
| Social Sciences | 6187 | 0.63 | 0.49 |
| Agricultural, Biological Science & Veterinary | 11975 | 0.63 | 0.48 |
| General (*Science, Nature, PNAS*) | 4520 | 0.63 | 0.49 |
| Biochemistry, Genetics & Molecular Biology | 17047 | 0.63 | 0.49 |
| Other Life & Health Sciences | 54555 | 0.62 | 0.50 |
| Health Professions & Nursing | 6068 | 0.61 | 0.50 |
| Medicine | 39265 | 0.61 | 0.51 |

**Table 5:** Summary of aspect-based analysis (all the numbers except the # of documents are percentages).

| QRR field | # of documents with at least 30 tweets | Title | Abstract | Methodology | Results & conclusions | Others removed |
|---|---|---|---|---|---|---|
| Other Life & Health Sciences | 54555 | 1.43 | 3.56 | 3.73 | 9.53 | 81.76 |
| Medicine | 39265 | 1.38 | 3.32 | 3.01 | 10.84 | 81.45 |
| Biochemistry, Genetics & Molecular Biology | 17047 | 1.56 | 3.02 | 4.88 | 7.13 | 83.40 |
| Agricultural, Biological Sciences & Veterinary | 11975 | 1.73 | 2.56 | 3.45 | 7.20 | 85.06 |
| Social Sciences | 6187 | 2.01 | 4.80 | 3.66 | 8.64 | 80.89 |
| Health Professions & Nursing | 6068 | 1.49 | 5.39 | 2.70 | 10.11 | 80.31 |
| General (*Science, Nature, PNAS*) | 4520 | 2.74 | 4.19 | 5.90 | 10.03 | 77.13 |
| Environmental Sciences | 2753 | 0.90 | 4.05 | 2.97 | 7.74 | 84.34 |
| Engineering | 2491 | 1.47 | 2.18 | 8.18 | 4.01 | 84.15 |
| Earth & Planetary Sciences | 1438 | 1.17 | 3.37 | 2.13 | 8.19 | 85.13 |
| Chemistry | 1414 | 1.40 | 2.79 | 8.66 | 3.70 | 83.45 |
| Physics & Astronomy | 1208 | 1.47 | 4.17 | 8.74 | 4.00 | 81.62 |
| Materials Sciences | 1094 | 1.09 | 1.90 | 12.04 | 1.90 | 83.08 |
| Computer Science | 979 | 1.59 | 4.98 | 7.36 | 6.57 | 79.50 |
| Mathematics | 950 | 1.14 | 2.79 | 7.64 | 5.68 | 82.77 |
| Economics, Business & Decision Sciences | 924 | 1.71 | 3.10 | 4.06 | 11.75 | 79.38 |
| Arts & Humanities | 468 | 2.29 | 5.20 | 3.74 | 7.07 | 81.70 |

## 4.5 Discussion

With the increased usage of the social media platforms for scholarly communications, altmetric data are of enhanced interest as it captures realtimerealtime scholarly communication data from online platforms and may be used as an early measure of the research impact. However, there is still a need to investigate tweeter data to analyse user sentiments relating to research articles in specific fields. Such a fine-grained investigation is required to fully utilize the findings of existing studies. We identify the sentiment of research communities with respect to their respective fields and to conduct an aspect-based analysis of user expressions related to their research articles. Such a fine-grained analysis may help research communities in improving their social media exchanges about the scientific articles to disseminate their scientific findings effectively and increase its impact.

We found that (i) Twitter usage in the domains of Mathematics, Engineering and Agriculture is inclined to the positive, while in Medicine and Environmental Sciences it tends towards the negative. Fields of research such as Chemistry and the Social Sciences were found overall to be neutral. Thus, research communities exhibit dissimilar sentiment towards their respective fields; and (ii) most tweets discuss research articles as a whole document; however, we saw a significant number where a specific aspect was discussed. Positive sentiment in tweets was observed to be more likely than negative. While the neutral sentiment is normally dominant in the whole-topic discussion, in aspect-based sentiment analysis it is almost matched by other sentiment expressions. This shows that the Twitter user is inclined to be specific in his or her opinion when discussing the aspects of an article.

# 5. Concluding remarks

We design harmonic means-based statistical measures to generate a specialized lexicon to conduct this investigation which helps improve the performance of the sentiment analysis task. This is because the general sentiment lexicons calculate the sentiment tendency of a word without considering domain knowledge. However, the sentiment contained in just a few words is inevitably domain-dependent. Therefore, generic sentiment lexicons used by SentiStrength report poor performance in various applications. Based on our newly generated lexicon, we designed a threshold-based mechanism to compute domain-wise article-level sentiment. Specifically, document-level sentiment analysis was performed to give a combined score for all tweets about a single altmetrics article. Each article was then given a score for positive and negative sentiment. This sentiment-analysis approach generated a ranking of altmetrics documents by this single sentiment score. The various fields of research were explored to ascertain the intent and behaviour of researchers and scholars. The results showed that Twitter usage in the domains of Mathematics, Engineering and Agriculture is inclined to the positive, while in Medicine and Environmental Sciences it tends towards the negative. Fields of research such as Chemistry and the Social

**Corresponding author:** saeed-ul-hassan@itu.edu.pk

Sciences were found overall to be neutral. Thus, research communities exhibit dissimilar sentiment towards their respective fields.

Document-level sentiment analysis was used to establish any correlation between sentiment score and citation score. For this purpose, the documents were allocated to three bins on the basis of their score: above 0.85; above 0.8; and above 0.75. Using correlation analysis, we found that highly positive documents, those scoring over 0.85, showed a moderate correlation to citation score. This suggests that positive sentiment in a tweet about a research article does indeed predict the article's popularity and has some relationship to it receiving somewhat more citations.

We also design a method to perform an aspect-based analysis of user expressions related to the research article, such as its title, abstract, methodology, conclusion and results. Various aspects of research articles were explored to examine which parts are commented upon by researchers in tweets. The results show that most tweets discuss research articles as a whole document; however, we saw a significant number where a specific aspect was discussed. Positive sentiment in tweets was observed to be more likely than negative. While the neutral sentiment is normally dominant in whole-topic discussion, in aspect-based sentiment analysis it is almost matched by other sentiment expressions. This shows that the Twitter user is inclined to be specific in his or her opinion when discussing the aspects of an article.

## 5.1 Implications

Research impact has used citation as the main indicator of research's standing, however, it takes years to see any measurable impact. On the other hand, researchers are increasingly going online to find and share information about science, as well as, they have been urged to consider how they can use social media platforms to engage with each other. With the increased usage of the social media platforms for scholarly communications, altmetric data are of enhanced interest as it captures realtime scholarly communication data from online platforms (e,g, Twitter) and may be used as an early measure of the research impact. Specifically, the papers cited in positive and neutral tweets have a greater impact than those not cited or cited in a negative tweet. However, there is still a need to investigate tweeter data to analyse user sentiments relating to research articles in specific fields. Such a fine-grained investigation is required to fully utilize the findings of existing studies.

**Corresponding author:** saeed-ul-hassan@itu.edu.pk

As mentioned earlier that this article presents a quantitative study to exploit tweet data to analyse user sentiments relating to different aspects of research articles in specific fields. This study helps us to understand the sentiments of online social exchanges of the scientific community on scientific literature, specifically the sentiment of tweets, for better visibility and qualitative assessment of these interesting big data of altmetrics. We identify the sentiment of research communities with respect to their respective fields and to conduct an aspect-based analysis of user expressions related to their research articles. Such a fine-grained analysis may help research communities in improving their social media exchanges about the scientific articles to disseminate their scientific findings effectively and increase their impact.

## 5.2 Limitations and Future works.

While there is a significant increase in Twitter usage in order to share research articles, the expression of opinion is still dominated by neutral sentiment, and the trends suggest no increase in sentiment expression. In further work that is undertaken over a longer duration, the Twitter mentions of a research article could be explored. Also, as we created a ranking of altmetrics articles on the basis of their Twitter popularity, we could follow up to see whether the topmost articles indeed attract a higher citation count, in time. Furthermore, a social-media campaign style can be detected in tweets about scholarly articles, as many are evidently a simple retweet, and this creates a great deal of duplication. Research can be carried out to establish the significance of retweets in terms of any correlation with citation. In addition, in terms of scoring documents, we believe that the influence of a Twitter user is significant; that is, the sentiment score of a tweet from a particularly relevant user should be heavily weighted. While aspect-based sentiment analysis was unable to capture a wide range of data, the aspects can be derived intellectually to increase the significance of these results. Moreover, less-good articles are sometimes used as a negative example in an article's literature review, thus future work could be undertaken on analysing the sentiment in a tweet in relation to a citation's opinion towards a scientific publication.

**Acknowledgments**

The authors (Salem Alelyani and Saeed-Ul Hassan) are grateful for the financial support received from King Khalid University for this research Under Grant No. R.G.P2/100/41.

Corresponding author: saeed-ul-hassan@itu.edu.pk

**Corresponding author:** saeed-ul-hassan@itu.edu.pk

## Appendix A

**Table A1:** Top 50 positive lexicon terms, with their positive and negative scores

| Token | Positive score (pos cdf hmean) | Negative score (neg cdf hmean) | Token | Positive score (pos cdf hmean) | Negative score (neg cdf hmean) |
|---|---|---|---|---|---|
| Excellent | 0.907660308 | 0.253328778 | Improves | 0.890779932 | 0.304454795 |
| Novel | 0.906440324 | 0.2557995 | Promising | 0.889829053 | 0.288539616 |
| Amazing | 0.905406722 | 0.257757634 | Improved | 0.888990591 | 0.304827953 |
| Congratulations | 0.905110313 | 0.248694131 | Worth | 0.888006505 | 0.302092593 |
| Awesome | 0.903971472 | 0.26272544 | Best | 0.886228491 | 0.322466558 |
| Success | 0.903082892 | 0.260565937 | Useful | 0.885983593 | 0.317457826 |
| Wow | 0.903005407 | 0.264781511 | Successful | 0.885539703 | 0.272322083 |
| Interested | 0.902443229 | 0.267758917 | Super | 0.884806715 | 0.271129548 |
| Exciting | 0.902263317 | 0.262783203 | Excited | 0.883732653 | 0.254014497 |
| Great | 0.901346915 | 0.295209104 | Neat | 0.88328853 | 0.269576177 |
| Positive | 0.900758366 | 0.273718857 | Fitness | 0.839063989 | 0.309667926 |
| Nice | 0.900716826 | 0.288556025 | Plays | 0.838340923 | 0.271447038 |
| Thanks | 0.900603947 | 0.276785536 | Save | 0.837633457 | 0.343615901 |
| Special | 0.899726703 | 0.273907706 | Stuff | 0.836890133 | 0.395109742 |
| Cool | 0.899106111 | 0.288502052 | Kind | 0.835780667 | 0.310959151 |
| Interesting | 0.898675388 | 0.308576729 | Very | 0.83485292 | 0.441671495 |
| Greater | 0.896300338 | 0.285598025 | Fun | 0.833788337 | 0.309398554 |
| Hope | 0.895501791 | 0.277879903 | Helpful | 0.830742627 | 0.357392534 |
| Love | 0.894677105 | 0.289159208 | Superior | 0.829893382 | 0.343208629 |
| Pretty | 0.893660856 | 0.28025927 | Colleagues | 0.829757063 | 0.407013242 |
| Improving | 0.893634078 | 0.285442743 | Wonderful | 0.829151154 | 0.254216552 |
| Please | 0.89359334 | 0.286969335 | Overview | 0.828697218 | 0.325720107 |
| Huge | 0.89281883 | 0.279777332 | Improvements | 0.827931768 | 0.300823749 |
| Interest | 0.892702527 | 0.288727767 | Winner | 0.826416998 | 0.262400833 |
| Fascinating | 0.891111422 | 0.300504873 | Article | 0.826234942 | 0.45688716 |

Corresponding author: saeed-ul-hassan@itu.edu.pk

**Table A1:** Top 50 negative lexicon terms, with their positive and negative scores

| Token | Negative score (neg cdf hmean) | Positive score (pos cdf hmean) | Token | Negative score (neg cdf hmean) | Positive ccore (pos cdf hmean) |
|---|---|---|---|---|---|
| Depression | 0.956523417 | 0.142962291 | Sad | 0.933212243 | 0.146280421 |
| Failure | 0.955566946 | 0.145001895 | Decrease | 0.931970753 | 0.209969222 |
| Chronic | 0.954612919 | 0.148733306 | Threat | 0.931410764 | 0.154818844 |
| Anxiety | 0.954035904 | 0.1451852 | Sorry | 0.930641447 | 0.176863056 |
| Loss | 0.95376771 | 0.151883033 | Problem | 0.930362917 | 0.217479084 |
| Worse | 0.953155776 | 0.152592792 | Complications | 0.929943039 | 0.16253538 |
| Fight | 0.951942612 | 0.149667313 | Difficult | 0.929870572 | 0.182274957 |
| Poor | 0.948370281 | 0.167972753 | Fail | 0.92825984 | 0.149107358 |
| Obesity | 0.947958065 | 0.169641955 | Challenge | 0.927242281 | 0.195327438 |
| Abuse | 0.945487713 | 0.143557614 | Harm | 0.924565382 | 0.214261433 |
| Critical | 0.945486351 | 0.175525367 | Cross | 0.922892138 | 0.191351394 |
| Decline | 0.944649103 | 0.173960254 | Violence | 0.922121721 | 0.144070395 |
| Risks | 0.943912604 | 0.179202624 | Bad | 0.921307242 | 0.244515638 |
| Low | 0.943589836 | 0.187245872 | Aggressive | 0.919158237 | 0.146882486 |
| Lack | 0.943111655 | 0.181049788 | Weak | 0.867458012 | 0.189337168 |
| Source | 0.94305159 | 0.181707386 | Harms | 0.866356927 | 0.155134184 |
| Risk | 0.943020831 | 0.19384778 | Depressed | 0.865741126 | 0.151294593 |
| Stress | 0.942724219 | 0.182659655 | Factor | 0.864080846 | 0.359182693 |
| Missed | 0.94139016 | 0.183218571 | Regardless | 0.863985589 | 0.174197141 |
| Problems | 0.940341809 | 0.190235998 | Complicated | 0.863504401 | 0.195533734 |
| Wrong | 0.939898407 | 0.186005692 | Inequality | 0.863401924 | 0.18313397 |
| Dependent | 0.939867264 | 0.175491201 | Beware | 0.862710192 | 0.169002881 |
| Obese | 0.939746991 | 0.185058127 | Controversial | 0.862344075 | 0.183446469 |
| Challenges | 0.934442513 | 0.203175391 | Fighting | 0.862208247 | 0.144446439 |
| Crisis | 0.934260093 | 0.147841295 | Waste | 0.861882475 | 0.200116067 |